\documentclass[twocolumn, amsmath,amssymb, superscriptaddress]{revtex4}

\usepackage{dcolumn}
\usepackage{bm}
\usepackage{amsmath}
\usepackage{amsfonts}
\usepackage{graphics}
\usepackage[pdftex]{graphicx}
\usepackage{times}
\usepackage{verbatim}
\usepackage[raggedright]{titlesec}
\usepackage[normalem]{ulem}
\begin{document}

\title{Languages cool as they expand: Allometric scaling and the decreasing need for new words}

\author{Alexander M. Petersen}
\email{alexander.petersen@imtlucca.it}
\affiliation{Laboratory for the Analysis of Complex Economic Systems, IMT Lucca Institute for Advanced Studies, Lucca 55100, Italy}
\author{Joel N. Tenenbaum}
\affiliation{Center for Polymer Studies and Department of Physics, Boston University, Boston, Massachusetts 02215, USA}
\affiliation{Operations and Technology Management, School of Management, Boston University, Boston, Massachusetts 02215, USA}
\author{Shlomo Havlin}
\affiliation{Minerva Center and Department of Physics, Bar-Ilan University, Ramat-Gan 52900, Israel}
\author{H. Eugene Stanley}
\affiliation{Center for Polymer Studies and Department of Physics, Boston University, Boston, Massachusetts 02215, USA}
\author{Matja{\v z} Perc}
\email{matjaz.perc@uni-mb.si}
\affiliation{Department of Physics, Faculty of Natural Sciences and
Mathematics, University of Maribor, Koro{\v s}ka cesta 160, SI-2000 Maribor, Slovenia}

\begin{abstract}
We analyze the occurrence frequencies of over 15 million words recorded in millions of books published during the past two centuries in seven different languages. For all languages and chronological subsets of the data we confirm that two scaling regimes characterize the word frequency distributions, with only the more common words obeying the classic Zipf law. Using corpora of unprecedented size, we test the allometric scaling relation between the corpus size and the vocabulary size of growing languages to demonstrate a decreasing marginal need for new words, a feature that is likely related to the underlying correlations between words. We calculate the annual growth fluctuations of word use which has a decreasing trend as the corpus size increases, indicating a slowdown in linguistic evolution  following language expansion.  This ``cooling pattern'' forms the basis of a third statistical regularity, which unlike the Zipf and the Heaps law, is dynamical in nature.
\end{abstract}

\maketitle

Books in libraries and  attics around the world
constitute an immense ``crowd-sourced'' historical record that traces the evolution of culture
  back beyond the limits of oral history.
However, the disaggregation of written language into individual books makes the longitudinal analysis of language a difficult open problem.
To this end, the book digitization project at {\it Google} Inc. presents a monumental step forward providing an enormous, publicly accessible, collection of written
language  in the form of the \textit{Google Books Ngram
  Viewer} web application \cite{google}.  Approximately 4\% of all books ever
published have been scanned, making available over $10^{7}$
occurrence time series (word-use trajectories) that archive cultural dynamics in seven
different languages over a period of more than two centuries.
This dataset highlights the utility of open ``Big Data,'' which is the gateway to ``metaknowledge'' \cite{evans_s11}, the
knowledge about knowledge. A digital data deluge is sustaining extensive interdisciplinary research efforts
towards quantitative insights into the social and natural sciences
\cite{ComplexitySociety,InnovAcc,lazer_s09,barabasi_np12,vespignani_np12}.

``Culturomics,''  the use of high-throughput data for the purpose of studying human culture, is a promising new empirical platform for gaining insight into subjects ranging from political history to epidemiology \cite{michel_s11}.
As first demonstrated by Michel et al. \cite{michel_s11}, the {\it Google} n-gram dataset is well-suited for examining the microscopic properties of an entire language ecosystem. Using this dataset to analyze the growth patterns of individual word frequencies, Petersen et al.~\cite{petersen_sr12} recently identified tipping points in the life trajectory of new words, statistical patterns that govern the fluctuations in word use, and quantitative measures for cultural memory.  The statistical properties of cultural memory, derived from the quantitative analysis of individual word-use trajectories, were also
investigated by Gao et al.~\cite{gao_jrsi12}, who found that words
describing social phenomena tend to have different long-range
correlations than words describing natural phenomena.

Here we study the growth and evolution of written language  by analyzing the macroscopic scaling patterns that characterize word-use.
Using the {\it Google} 1-gram data collected at the 1-year time resolution over the period 1800-2008, we quantify the annual fluctuation scale of words
within a given corpora and show that
languages can be said to ``cool by expansion.''
This effect constitutes a dynamic law, in contrast to the static laws of Zipf and Heaps which
are founded upon snapshots of single texts.
The Zipf law
\cite{zipf_49,tsonis_c97,serrano_pone09,cancho_jql01,cancho_epjb05,cancho_pnas03,baek_njp11},
quantifying the distribution of word frequencies, and the Heaps law
\cite{heaps_1978,serrano_pone09,bernhardsson_njp09,bernhardsson_pa11},
relating the size of a corpus to the vocabulary size of that corpus, are classic paradigms that capture many complexities of language
in remarkably simple statistical patterns. While these
laws have been exhaustively tested on relatively small snapshots of
empirical data, here we test the
validity of these laws using extremely large corpora.

Interestingly, we observe two scaling regimes in the probability density
functions of word usage, with the Zipf law holding only for the set of more
frequently used words, referred to as the ``kernel lexicon'' by Ferrer i Cancho et al.
\cite{cancho_jql01}. The word frequency distribution for the rarely used words constituting the ``unlimited lexicon'' \cite{cancho_jql01} obeys a distinct scaling law, suggesting that rare words belong to a
distinct class.  This ``unlimited lexicon'' is populated by highly technical words, new words, numbers,  spelling variants of kernel words, and optical character recognition (OCR) errors.

Many new words start in relative obscurity, and their eventual importance can be under-appreciated by their initial frequency.
This fact is closely related to the information cost of introducing new words and concepts.
For single topical texts, Heaps observed that the  vocabulary
size exhibits sub-linear growth with document size \cite{heaps_1978}. Extending this concept to entire corpora, we find a  scaling relation that indicates a decreasing ``marginal need'' for
new words which are the manifestation of cultural evolution and the seeds for language growth.
We introduce a pruning method to study the role of infrequent words on the allometric scaling properties of language.
By studying progressively smaller sets of the kernel lexicon we can better understand the marginal utility of the core words.
The pattern that arises for all languages analyzed provides insight into the intrinsic  dependency structure between words.

The correlations in word use can also be author and topic dependent.   Bernhardsson et al. recently introduced  the ``metabook'' concept  \cite{bernhardsson_njp09,
  bernhardsson_pa11}, according to which word-frequency structures are
author-specific: the word-frequency characteristics of a
random excerpt from a compilation of everything that a specific author
could ever conceivably write (his/her ``metabook'') should accurately
match those of the author's actual writings.   It is not immediately obvious  whether a
compilation of all the metabooks of all authors would still conform to the Zipf law and
the Heaps law.  The immense size and time span of the {\it Google} n-gram
dataset allows us to examine this question in detail.

\section*{Results}

\begin{figure}
\centering{\includegraphics[width = 8.5cm]{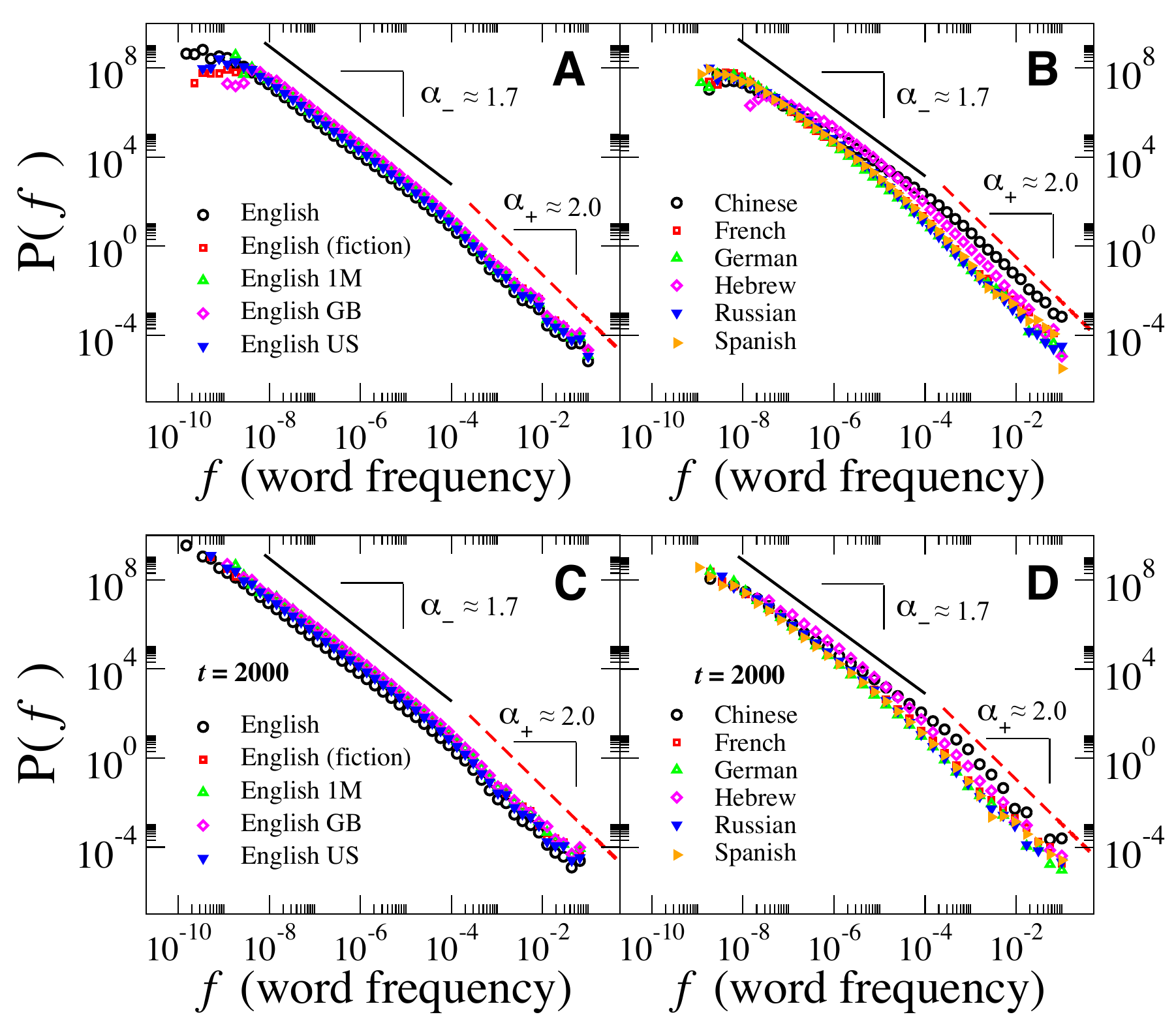}}
\caption{Two-regime scaling distribution of word
  frequency. The kink in the probability density functions $P(f)$ occurs
  around $f_{\times} \approx 10^{-5}$ for each corpora analyzed (see
  legend).  (A,B) Data from all years are aggregated into a single
  distribution.  (C,D) $P(f)$ comprising data from only year $t=2000$
  providing evidence that the distribution is stable even over shorter
  time frames and likely emerges in corpora that are sufficiently large
  to be comprehensive of the language studied.   For details concerning
  the scaling exponents we refer to Table~\ref{TableSummary1} and the
  main text.}
\label{fpdfall}
\end{figure}

\noindent{\bf Longitudinal analysis of written language.} Allometric scaling analysis \cite{kleiber_32} is used to quantify the role of
system size on general phenomena characterizing a system, and has been applied to systems as diverse as the
metabolic rate of mitochondria \cite{west_pnas02} and city growth
\cite{makse_n95, makse_pre98, Rozenfeld_pnas08, gabaix_qje99, bettencourt_pnas07, batty_s08, rozenfeld_aer11}.
Indeed, city growth shares two common features with the growth of
written text: (i) the Zipf law is able to describe the distribution of
city sizes regardless of country or the time period of the data
\cite{gabaix_qje99}, and (ii) city growth has inherent constraints due
to geography, changing labor markets and their effects on opportunities
for innovation and wealth creation \cite{bettencourt_pnas07,batty_s08},
just as vocabulary growth is constrained by human brain capacity and the
varying utilities of new words across users \cite{cancho_jql01}.

We construct a word counting framework by first defining the quantity
$u_{i}(t)$ as the number of times word $i$ is used in year $t$.  Since
the number of books and the number of distinct words grow dramatically
over time, we define the {\it relative} word use, $f_{i}(t)$, as the
fraction of the total body of text occupied by word $i$ in the same year
\begin{equation}
f_{i}(t) \equiv u_{i}(t)/N_{u}(t),
\end{equation}
where the quantity $N_{u}(t) \equiv \sum_{i=1}^{N_{w}(t)} u_{i}(t)$ is
the total number of indistinct word uses while $N_{w}(t)$ is the total number of distinct words digitized from books printed in year $t$.  Both the $N_{w}$  (``types'' giving the vocabulary size) and the $N_{u}$ (``tokens'' giving  the size of the body of
text) are generally increasing
over time.\\

\begin{figure}
\centering{\includegraphics[width = 8.5cm]{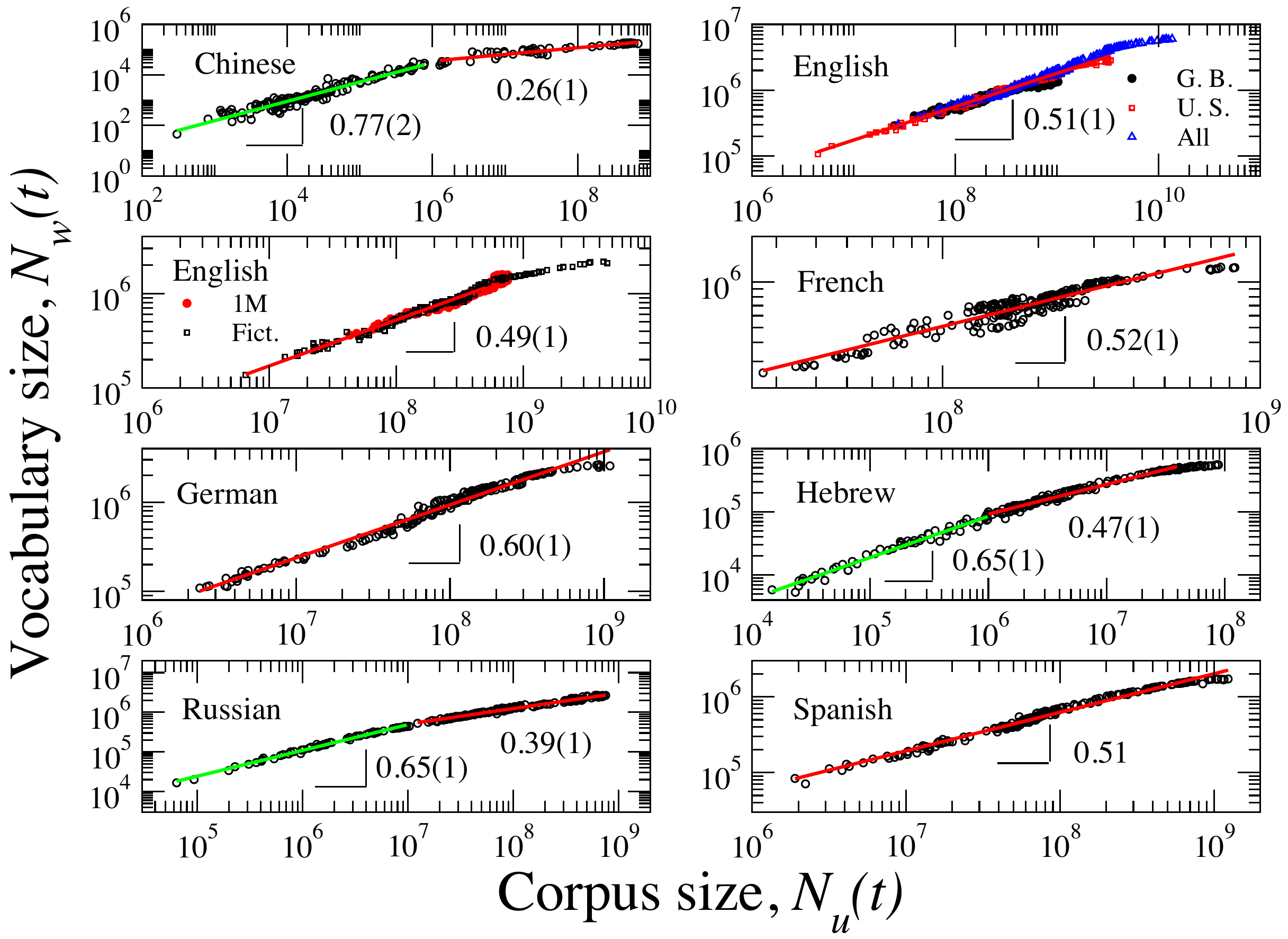}}
\caption{Allometric scaling of language. Scatter plots of the output corpora size $N_{u}$ given the
  empirical vocabulary size $N_{w}$ using all data ($U_{c}=0$) over the 209-year period
  1800--2008. Shown are OLS estimation of the  exponent $b$ quantifying the Heaps' law relation $N_{w}\sim [N_{u}]^{b}$. }
\label{HeapsLawfc}
\end{figure}

\noindent{\bf The Zipf law and the two scaling regimes.}
Zipf investigated a number of bodies of literature and observed that the
frequency of any given word is roughly inversely proportional to its
rank \cite{zipf_49}, with the frequency of the $z$-ranked word given by
the relation
\begin{equation}
f(z) \sim z^{-\zeta},
\label{zipfr}
\end{equation}
with a scaling exponent $\zeta \approx 1$.  This empirical law has been
confirmed for a broad range of data, ranging from income rankings, city
populations, and the varying sizes of avalanches, forest fires
\cite{newman_cp05} and firm size \cite{stanley_EL95}
 to the linguistic features of nonconding DNA \cite{Mantegna_PRE95}.
  The Zipf law can be derived through the
``principle of least effort,'' which minimizes the communication noise
between speakers (writers) and listeners (readers)
\cite{cancho_pnas03}. The Zipf law has been found to hold for a large
dataset of English text \cite{cancho_jql01}, but there are interesting
deviations observed in the lexicon of individuals diagnosed with schizophrenia  \cite{cancho_epjb05}.  Here, we
also find statistical regularity in the distribution of relative word
use for 11 different datasets, each comprising more than half a million
distinct words taken from millions of books \cite{michel_s11}.

\begin{figure}
\centering{\includegraphics[width=8.5cm]{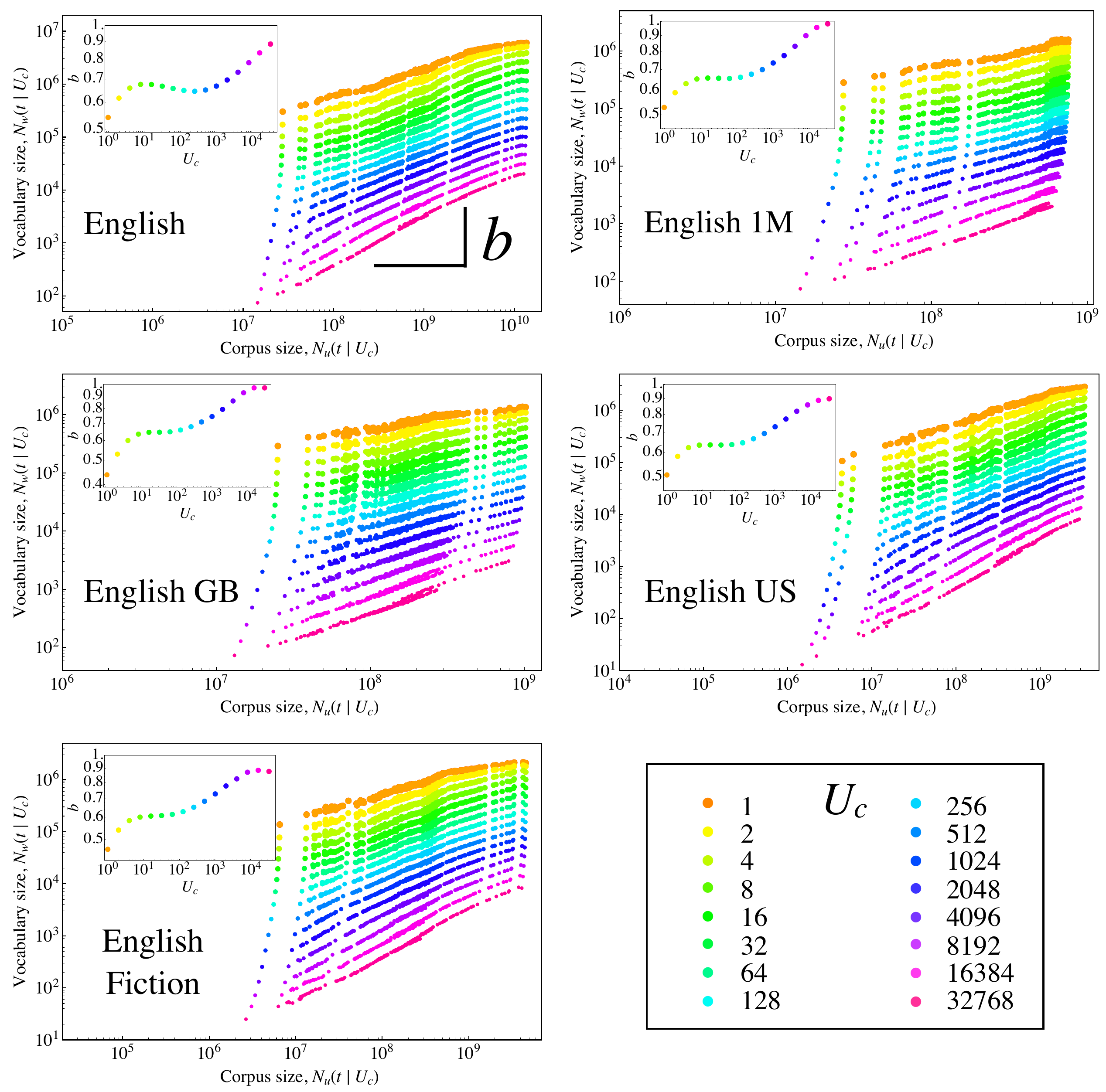}}
  \caption{Pruning reveals the variable marginal return of  words. The Heaps scaling exponent $b$ depends on the extent of the inclusion of the rarest words. For a given corpora and $U_{c}$ value  we make a scatter plot  between $N_{w}(t|U_{c})$ and $N_{u}(t|U_{c})$
     using
     words with $u_{i}(t) \geq U_{c}$. (Panel Inset)  We  use OLS estimation to estimate the scaling exponent $b(U_{c})$ for the model $N_{w}(t|U_{c})\sim [N_{u}(t|U_{c})]^{b}$
  to show that $b(U_{c})$ increases from approximately $0.5$ towards unity as we prune the corpora of extremely rare words.
   Our longitudinal language analysis provides insight into  the structural importance of the most frequent words which are used more times per appearance and which play a crucial role in the usage of new and rare words.}
\label{MultifcEng}
\end{figure}

\begin{figure}
\centering{\includegraphics[width=8.5cm]{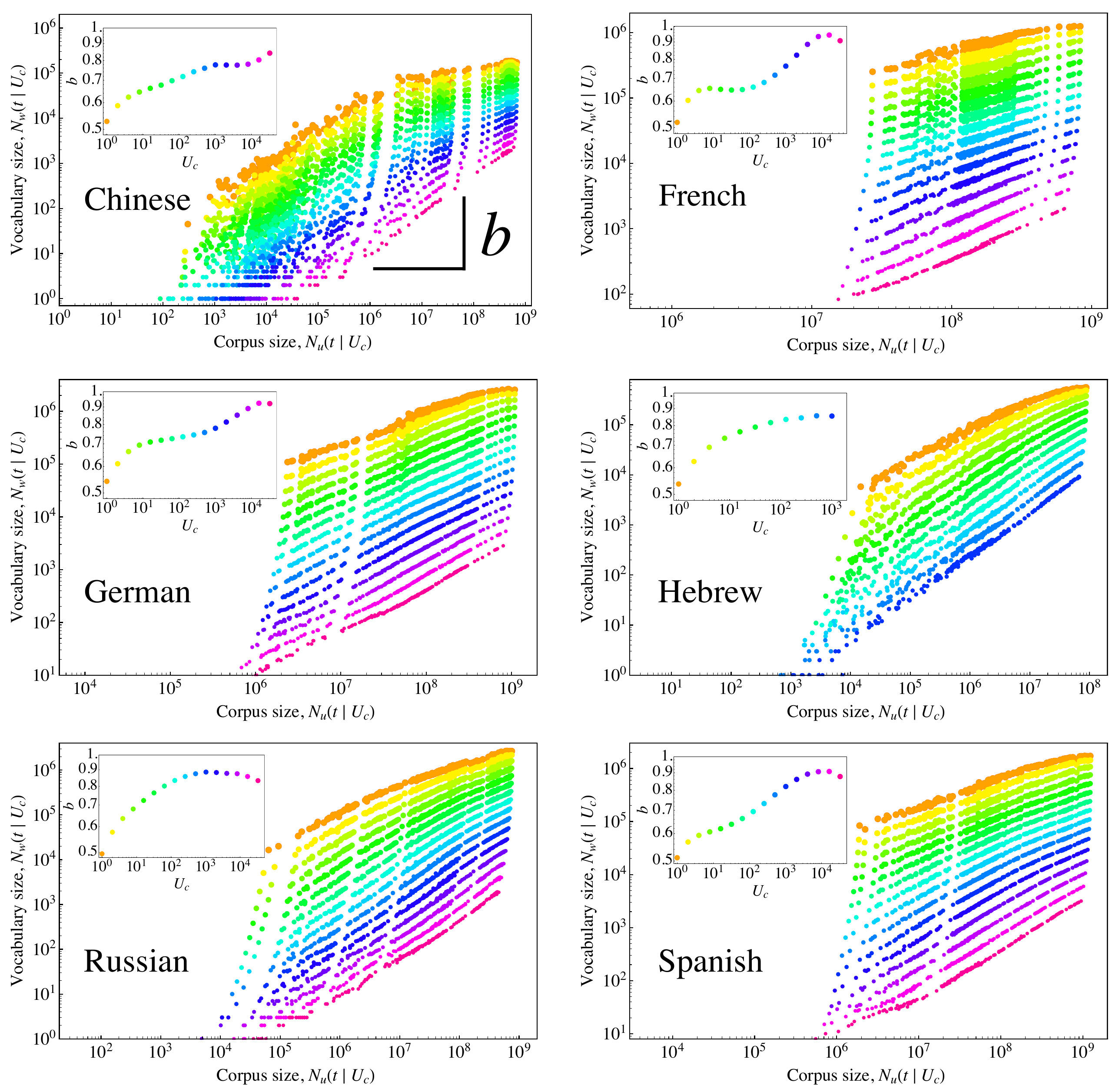}}
  \caption{ Pruning reveals the variable marginal return of  words.  The Heaps scaling exponent $b$ depends
  on the extent of the inclusion of the rarest words. For a given corpora and $U_{c}$ value  we make a scatter plot  between $N_{w}(t|U_{c})$ and $N_{u}(t|U_{c})$    using     words with $u_{i}(t) \geq U_{c}$, using the same data color-$U_{c}$ correspondence as in Fig. \ref{MultifcEng}. (Panel Inset) We  use OLS estimation to estimate the scaling exponent $b(U_{c})$ for the model $N_{w}(t|U_{c})\sim [N_{u}(t|U_{c})]^{b}$
  to show that $b(U_{c})$ increases from approximately $0.5$ towards unity as we prune the corpora of extremely rare words.
    Our longitudinal language analysis provides insight into  the structural importance of the most frequent words which are used more times per appearance and which play a crucial role in the usage of new and rare words.}
\label{MultifcOther}
\end{figure}

\begin{table*}
\caption{   Summary of the scaling exponents characterizing the Zipf law
and the Heaps law.  To calculate $\sigma_{r}(t|f_{c})$ (see
  Figs.~\ref{YSigma} and \ref{CorpusTotalSizefcSigmaScaling}) we use
  only the relatively common words that meet the criterion that their
  average word use $\langle f_{i} \rangle$ over the entire word history
  is larger than a threshold $f_{c}=10/Min[N_{u})(t)]$ listed in the first column
  for each corpus.  The $b$ values shown are calculated using all words
  ($U_{c}=0$). The ``unlimited lexicon'' scaling exponent
  $\alpha_{-}(t)$ is calculated for $10^{-8} < f <10^{-6}$ and the
  ``kernel lexicon'' exponent $\alpha_{+}(t)$ is calculated for $10^{-4}
  < f <10^{-1}$ using the maximum likelihood estimator method for each
  year.  The average and standard deviation $(\langle \dotsm \rangle \pm
  \sigma)$ listed are computed using the $\alpha_{+}(t)$ and $
  \alpha_{-}(t)$ values over the 209-year period 1800--2008 (except for
  Chinese, which is calculated from 1950--2008 data).  We show the Zipf scaling exponent calculated as
  $\zeta = 1/\Big(\langle \alpha_{+} \rangle -1\Big)$. The last column indicates the $\beta$ scaling exponents from Fig. \ref{CorpusTotalSizefcSigmaScaling}(A). }
\bigskip
\begin{tabular}{@{\vrule height 10.5pt depth4pt  width0pt}lc|c|c|c|c|c|}
  \multicolumn6c{Scaling  parameters}\\
\noalign{
\vskip-11pt} Corpus\\
\cline{2-7}
\vrule depth 6pt width 0pt (1-grams)& $Min[N_{u}(t)]$ & $b(U_{c}=0) $ & $\langle
\alpha_{-}\rangle $ &  $\langle \alpha_{+} \rangle$  & $\zeta$ & $\beta$  \\
\hline
\hline
Chinese & $35,394$  & $0.77 \pm 0.02$ & 1.49 $\pm 0.15$  &
1.91 $\pm 0.04 $ &  1.10 $\pm 0.05 $ &$ 0.20\pm 0.01$\\
English & $42,786,702$  & $0.54\pm 0.01$ & 1.73 $\pm 0.05$  & 2.04
$\pm 0.06$ &  0.96 $\pm 0.06$ & $0.19\pm0.01 $ \\
English fiction & $13,184,111$   & $0.49 \pm 0.01$ & 1.68 $\pm
0.10$ & 1.97 $\pm 0.04$ &  1.03 $\pm 0.04$ & $0.18\pm 0.01$ \\
English GB & $38,956,621$  & $0.44 \pm 0.01$ & 1.71 $\pm 0.07$  &
2.02 $\pm 0.05$ &  0.98 $\pm 0.05$ & $0.17\pm 0.01$ \\
English US & $5,821,340$  & $0.51 \pm 0.01$ & 1.70 $\pm 0.08$  &
2.03 $\pm 0.06$ &  0.97 $\pm 0.06$ & $0.18\pm 0.01$ \\
English 1M & $42,778,968$  & $0.53 \pm 0.01$ & 1.71 $\pm 0.04$  &
2.04 $\pm 0.06$ &  0.96 $\pm 0.06$ & $0.25\pm 0.01$ \\
French & $34,198,362$  & $0.52 \pm 0.01$ & 1.69 $\pm 0.06$  & 1.98
$\pm 0.04$ & 1.02 $\pm 0.04$ & $0.26\pm 0.01$ \\
German & $2,274,842$  & $0.60 \pm 0.01$ & 1.63 $\pm 0.16$  & 2.02
$\pm 0.03$ &  0.98 $\pm 0.03$ & $0.27\pm 0.01$ \\
Hebrew & $9,482$ & $0.47 \pm 0.01$ & 1.34  $\pm 0.09$ & 2.06
$\pm 0.05$ & 0.94 $\pm 0.05$ & $ 0.35\pm 0.01$  \\
Russian & $6,944,366$ & $0.65\pm 0.01$ & 1.55  $\pm 0.17$ & 2.04
$\pm 0.06$ &   0.96 $\pm 0.06$ & $0.08\pm 0.01$ \\
Spanish & $1,777,563$  & $0.51 \pm 0.01$ & 1.61 $\pm 0.15$  &
2.07 $\pm 0.04$ &  0.93 $\pm 0.04$ & $0.26\pm 0.01 $ \\
\hline
\end{tabular}
\label{TableSummary1}
\end{table*}

Figure~\ref{fpdfall} shows the probability density functions $P(f)$
resulting from data aggregated over all the years (A,B) as well as over 1-year periods as demonstrated for
the year $t=2000$ (C,D).  Regardless of the language and the
considered time span, the probability density functions are
characterized by a striking two-regime scaling, which was first noted by
Ferrer i Cancho and Sol{\'e} \cite{cancho_jql01}, and can be quantified
as
\begin{equation}
P(f) \sim\begin{cases}
f^{-\alpha_-}, & \mbox{if } f< f_{\times} \mbox{ [``unlimited lexicon'']}\\
f^{-\alpha_+}, & \mbox{if } f> f_{\times} \mbox{ [``kernel lexicon'']} \ .
\end{cases}
\label{PDFf}
\end{equation}
These two regimes, designated ``kernel lexicon'' and ``unlimited
lexicon,'' are thought to reflect the cognitive constraints of the
brain's finite vocabulary \cite{cancho_jql01}.  The specialized words found in the unlimited lexicon
are not universally shared and are used significantly less frequently than the words in the kernel lexicon.  This is reflected in the kink in the probability density
functions and gives rise to the anomalous two-scaling distribution shown in
Fig.~\ref{fpdfall}.

The exponent $\alpha_{+}$ and the corresponding rank-frequency scaling
exponent $\zeta$ in Eq.~(\ref{zipfr})  are related asymptotically by \cite{cancho_jql01}
\begin{equation}
\alpha_{+} \approx 1 + 1/\zeta,
\label{zipfpdf}
\end{equation}
with no analogous relationship for the unlimited lexicon values $\alpha_{-}$ and $\zeta_{-}$.
Table~\ref{TableSummary1} lists the average $\alpha_{+}$ and
$\alpha_{-}$ values calculated by aggregating $\alpha_{\pm}$ values for
each year using a maximum likelihood estimator for the power-law
distribution \cite{powlawMLE}.  We characterize the two scaling regimes
using a crossover region around $f_{\times} \approx10^{-5}$ to
distinguish between $\alpha_{-}$ and $\alpha_{+}$: (i) $10^{-8}\leq f
\leq 10^{-6}$ corresponds to $\alpha_{-}$ and (ii) $10^{-4}\leq f \leq
10^{-1}$ corresponds to $\alpha_{+}$.  For the words
that satisfy $f \gtrsim f_{\times}$ that comprise
the kernel lexicon, we verify the Zipf scaling law $\zeta \approx 1$
(corresponding to $\alpha \approx 2$) for all corpora analyzed.  For the
unlimited lexicon regime $f \lesssim f_{\times}$, however, the Zipf law is
not obeyed, as we find $\alpha_{-} \approx 1.7$.  Note that $\alpha_{-}$
is significantly smaller in the Hebrew, Chinese, and the Russian
corpora, which suggests that a more generalized version of the Zipf
law \cite{cancho_jql01}  may be needed, one which is slightly language-dependent,
especially when taking into account the  usage of specialized words from
the unlimited lexicon.\\

\noindent{\bf The Heaps law and the increasing marginal returns of new words.}
Heaps observed that vocabulary size, i.e.~the number of distinct words,
exhibits a sub-linear growth with document size \cite{heaps_1978}.
This observation has important implications for
the ``return on investment'' of a new word as it is established and
becomes disseminated throughout the literature of a given language.  As a
proxy for this return, Heaps studied how often new words are invoked in
lieu of preexisting competitors and examined the linguistic value of new
words and ideas by analyzing the relation between the total number of
words printed in a body of text $N_{u}$, and the number of these which
are distinct $N_{w}$, i.e.~the vocabulary size \cite{heaps_1978}.  The
marginal returns of new words, $\partial N_{u}/\partial N_{w}$
quantifies the impact of the addition of a single word to the vocabulary
of a corpus on the aggregate output (corpus size).

For individual books, the empirically-observed scaling relation between
$N_u$ and $N_w$ obeys
\begin{equation}
N_{w}\sim (N_{u})^{b},
\label{HLaw}
\end{equation}
with $b<1$, with Eq.~(\ref{HLaw}) referred to as ``the Heaps law''.
It has subsequently been found that Heaps' law emerges naturally in systems that can be described as sampling from an underlying Zipf distribution.
In an information theoretic formulation of the
the abstract concept of  word cost, B. Mandelbrot predicted the relation $b=1/\zeta$ in 1961  \cite{Mandelbrot61},
where $\zeta$ is the scaling exponent corresponding to $\alpha_{+}$, as  in Eqs.~(\ref{PDFf}) and ~(\ref{zipfpdf}).
This prediction is limited to relatively small texts where the unlimited lexicon, which manifests in the $\alpha_{-}$ regime, does not play a significant role.
A  mathematical extension of this result for general underlying rank-distributions is also provided by Karlin \cite{Karlin67} using an infinite urn scheme, and extended to broader classes of heavy-tailed distributions recently by Gnedin et al. \cite{PowerLawOccupancyHeaps}.
Recent research efforts using stochastic master equation techniques to model the growth of a book have also predicted this intrinsic relation  between
Zipf's law and Heaps' law \cite{DerivHeaps,serrano_pone09,FiniteSizeHeaps}.

\begin{figure*}
\centering{\includegraphics[width = 18cm]{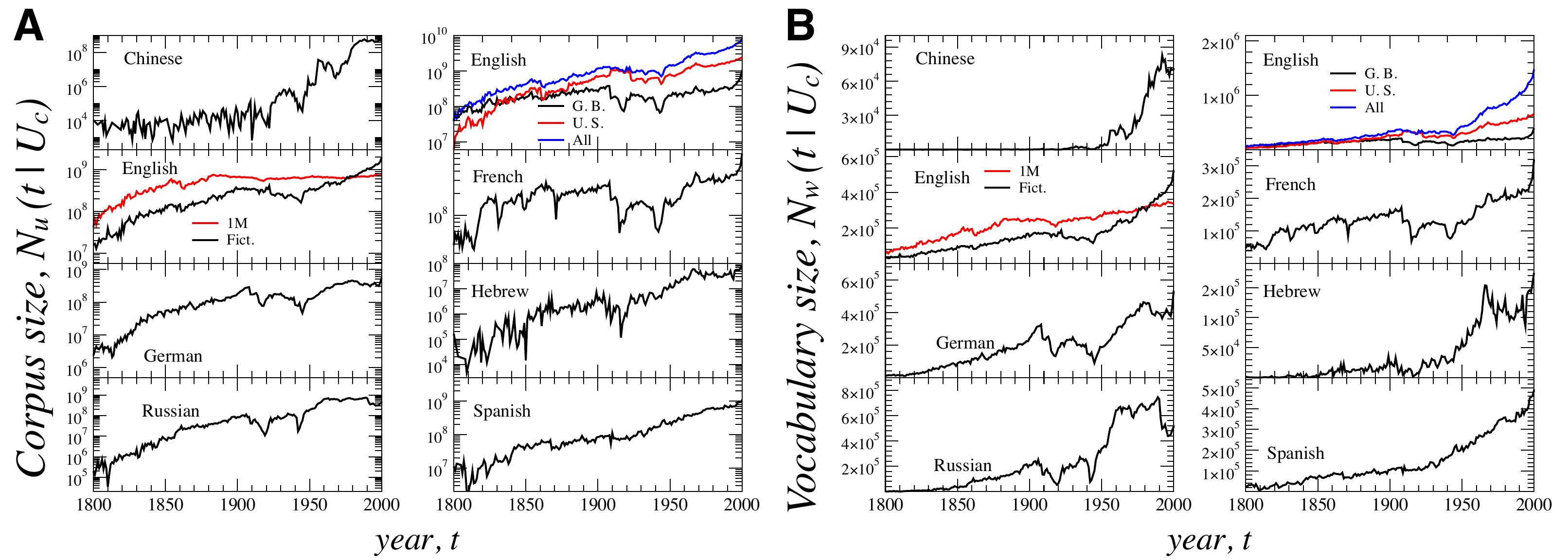}}
\caption{Literary productivity and
  vocabulary size in the {\it Google Inc.} 1-gram dataset over the past two
  centuries.  (A) Total size of the different corpora $N_{u}(t | U_{c})$
  over time, calculated by using words that satisfy $u_{i}(t) \geq U_{c} \equiv 16$ to eliminate extremely rare 1-grams.  (B) Size of
  the written vocabulary $N_{w}(t | U_{c})$ over time, calculated under
  the same conditions as (A).}
\label{VolumeCorpusTotalSizefc}
\end{figure*}

Figure~\ref{HeapsLawfc} confirms a sub-linear scaling ($b<1$) between
$N_u$ and $N_w$ for each corpora analyzed.
These results show how the marginal returns of new words are given by
\begin{equation}
\frac{\partial N_{u}}{\partial N_{w}} \sim (N_{w})^{(1-b)/b},
\label{marginal}
\end{equation}
which is an increasing function of $N_w$ for $b<1$.   Thus, the
relative increase in the induced volume of written languages is larger
for new words than for old words.  This is likely due to the fact that
new words are typically technical in nature, requiring additional
explanations that put the word into context with pre-existing words.
Specifically, a new word requires the additional use of
 preexisting words as a result of both (i) the explanation of the content of the new word
  using existing technical terms, and (ii) the grammatical
  infrastructure necessary for that explanation.  Hence, there are
  large spillovers in the size of the written corpus that follow from
  the intricate dependency structure of language stemming from the various
  grammatical roles  \cite{SemanticNetwork,markosova_pa08}.

\begin{figure}
\centering{\includegraphics[width = 8.5cm]{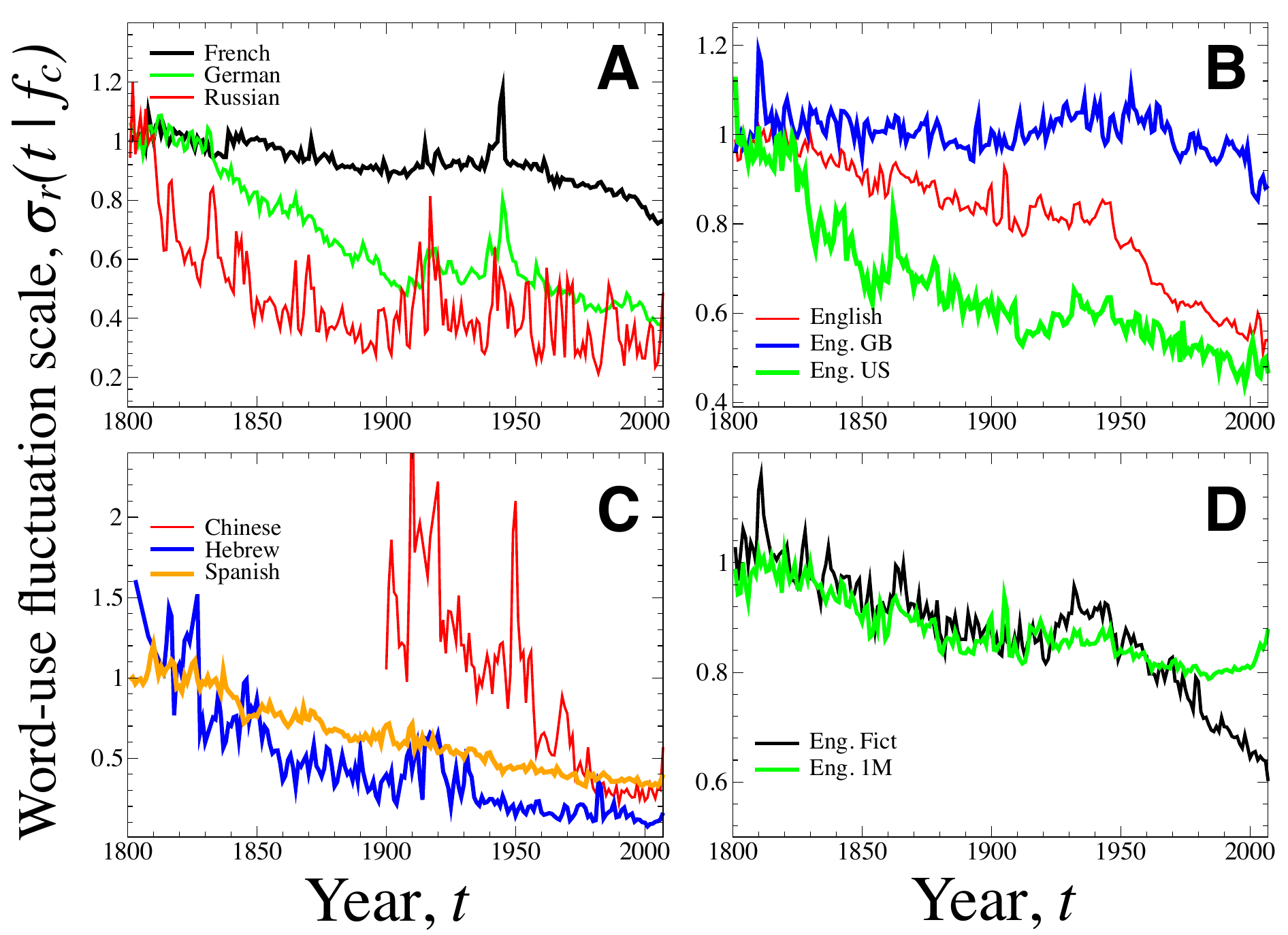}}
\caption{Non-stationarity in the characteristic growth
  fluctuation of word use.  The standard deviation $\sigma_{r}(t|f_{c})$ of the
  logarithmic growth rate $r_{i}(t)$ is presented for all examined
  corpora.  There is an overall decreasing trend arising from the
  increasing size of the corpora, as depicted in
  Fig.~\ref{VolumeCorpusTotalSizefc}(A).  On the other hand, the steady
  production of new words, as depicted in
  Fig.~\ref{VolumeCorpusTotalSizefc}(B) counteracts this effect.   We
  calculate $\sigma_{r}(t|f_{c})$ using the relatively common words that meet
  the criterion that their average word use $\langle f_{i} \rangle$ over
  the entire word history  $T_{i}$ (using words with lifetime $T_{i} \geq 10$ years) is larger than a threshold $f_{c} \equiv 1/Min[N_{u}(t)]$ (see
  Table~\ref{TableSummary1}).}
\label{YSigma}
\end{figure}

In order to investigate the role of rare and new words, we calculate $N_u$
and $N_w$ using only   words that have appeared at  least
$U_{c}$ times. We select the absolute number of uses as a word use threshold because
a word in a given year can not appear with a frequency less than $1/N_{u}$, hence any criteria
using relative frequency would necessarily introduce a bias for small corpora samples.
This choice also eliminates words that can spuriously arise from Optical Character Recognition (OCR)
errors in the digitization process and also  from intrinsic spelling
errors and orthographic spelling variations.

Figures~\ref{MultifcEng} and \ref{MultifcOther} show the relational dependence of $N_u$ and $N_w$
 on the exclusion of low-frequency words using a
variable cutoff $U_{c} =2^{n}$ with $n=0 \dotsc11$.   As $U_{c}$ increases the Heaps scaling exponent increases from $b\approx 0.5$, approaching $b \approx 1$, indicating that core words are structurally integrated into language as a proportional background.
Interestingly, Altmann et al. \cite{altmann_pone11} recently showed that ``word niche'' can be  an essential factor in modeling word use dynamics . New niche words,  though they are marginal increases to a language's lexicon, are themselves anything but ``marginal'' - they are
core words within a subset of the language. This is particularly the case in online communities in which
individuals strive to distinguish themselves on short timescales by
developing stylistic jargon,  highlighting how language patterns can be context dependent.

\begin{figure*}
\centering{\includegraphics[width = 17cm]{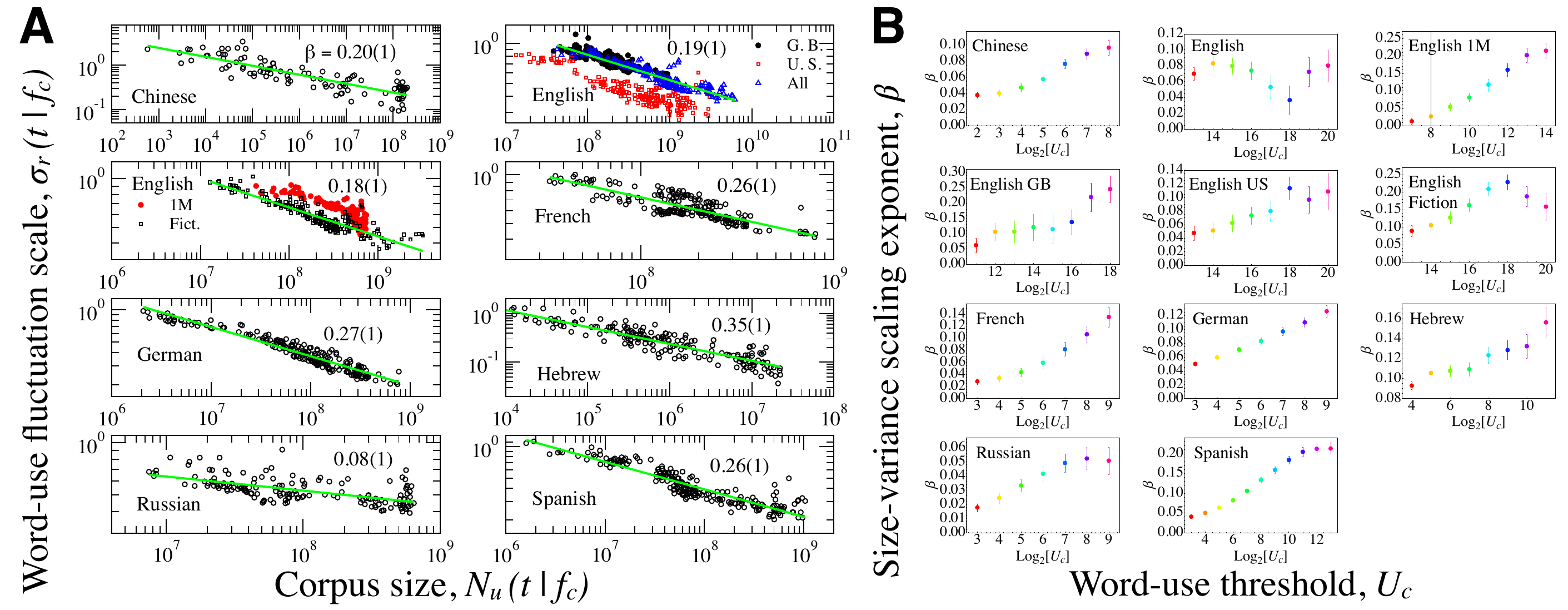}}
\caption{ Growth fluctuation of
  word use scale with the size of the corpora. (A) Depicted is the
  quantitative relation in Eq.(\ref{eqSV}) between $\sigma_{r}(t|f_{c})$
  and the corpus size $N_{u}(t|f_{c})$.
  We
  calculate $\sigma_{r}(t|f_{c})$ using the relatively common words that meet
  the criterion that their average word use $\langle f_{i} \rangle$ over
  the entire word history  (using  words with lifetime $T_{i} \geq 10$ years) is larger than a threshold $f_{c} \equiv 10/Min[N_{u}(t)]$  (see
  Table~\ref{TableSummary1}).
  We show
  the language-dependent scaling value $\beta \approx 0.08-0.35$ in each
  panel. For each language we show the value of the ordinary least squares best-fit  $\beta$ value with the standard error in parentheses.
 (B) Summary of $\beta(U_{c})$ exponents calculated using a use-threshold $U_{c}$, instead of a frequency threshold $f_{c}$ as used in (A). Error bars indicate the standard error in the OLS regression. We perform this additional analysis in order to provide alternative
 insight into the role of extremely rare words. For increasing $U_{c}$  the $\beta(U_{c})$ value for each corpora increases from $\beta \approx 0.05$ to  $\beta < 0.25$. This language pruning method quantifies the role of new rare words (also including OCR errors, spelling and other orthographic variants), which are the significant components of language volatility.
  }
\label{CorpusTotalSizefcSigmaScaling}
\end{figure*}

We now return to the relation between Heaps' law and Zipf's law.
Table~\ref{TableSummary1} summarizes the $b$ values calculated by means of ordinary least squares
regression using $U_{c} = 0$ to relate $N_{u}(t)$ to $N_{w}(t)$.
For $U_{c}=1$ we find that  $b \approx 0.5$ for all languages
analyzed, as expected from Heaps law, but for $U_{c} \gtrsim 8$ the $b$ value significantly deviates from $0.5$, and for  $U_{c} \gtrsim 1000$ the $b$ value
begins to saturate approaching unity.
Considering that  $\alpha_{+} \approx 2$ implies $\zeta \approx 1$
for all corpora, Figures~\ref{MultifcEng} and \ref{MultifcOther} shows that we can
confirm the relation $b(U_{c}) \approx 1/\zeta$ only for the more pruned corpora that require relatively large $U_{c}$.
This hidden feature of the  scaling relation highlights the underlying
structure of language, which forms a dependency network between the
common words of the kernel lexicon and their more esoteric counterparts in the unlimited lexicon.   Moreover, the  function
$\partial N_{w}/\partial N_{u}\sim (N_{u})^{b-1}$ is a monotonically decreasing function for  $b<1$, demonstrating the
{\it decreasing marginal need} for additional words as a corpora grows. In other words, since we get more
and more ``mileage'' out of new words in an already large language,
additional words are needed less and less.\\

\noindent{\bf Corpora size and word-use fluctuations.}
Lastly, it is instructive to examine how vocabulary size $N_{w}$ and the overall size of the corpora $N_{u}$ affect fluctuations in word use.
Figure~\ref{VolumeCorpusTotalSizefc} shows how $N_{w}(t)$ and $N_{u}(t)$ vary
over time over the past two centuries.  Note that, apart from the periods
during the two World Wars, the number of words printed, which we will
refer to as the ``literary productivity'', has been increasing over
time.  The number of distinct words (vocabulary size) has also increased reflecting basic social and technological advancement \cite{michel_s11}.

To investigate the role of fluctuations, we focus on the logarithmic growth rate, commonly used in finance and economics
\begin{eqnarray}
r_{i}(t) &\equiv& \ln f_{i}(t+\Delta t)-\ln f_{i}(t) = \ln \Big[ \frac{f_{i}(t+\Delta t)}{f_{i}(t)}\Big],
\label{r2}
\end{eqnarray}
to measure the relative growth of word use over 1-year periods, $\Delta
t \equiv$ 1 year.   Recent quantitative analysis on the distribution
$P(r)$ of word use growth rates $r_{i}(t)$ indicates that annual fluctuations in word use deviates significantly from the predictions of null
models for language evolution \cite{petersen_sr12}.

We define an aggregate fluctuation scale, $\sigma_{r}(t|f_{c})$, using a frequency cutoff $f_{c}\propto 1/Min[N_{u}(t)]$ to eliminate infrequently used words. The quantity $Min[N_{u}(t)]$ is the minimum corpora size over the period of analysis, and so $1/Min[N_{u}(t)]$ is an upper bound for the minimum observed frequency for words in the corpora.
Figure~\ref{YSigma} shows $\sigma_{r}(t|f_{c})$, the standard deviation of
$r_{i}(t)$ calculated across all words that  satisfy the condition $\langle f_{i} \rangle \geq f_{c}$ for words with lifetime $T_{i} \geq 10$ years, using  $f_{c}=1/Min[N_{u}(t)]$.
  Visual inspection suggests a general decrease in
$\sigma_{r}(t|f_{c})$ over time, marked by sudden increases during times of
political conflict.   Hence, the persistent increase in the volume of
written language is correlated with a persistent downward trend what could be thought of
as the ``system temperature''
$\sigma_{r}(t|f_{c})$: as a language grows and matures it also ``cools off''.

Since this cooling   pattern could arise as a simple artifact of an independent identically distributed (i.i.d) sampling from an increasingly large  dataset,
we test the scaling of  $\sigma_{r}(t|f_{c})$ with corpora size.
Figure~\ref{CorpusTotalSizefcSigmaScaling}(A) shows that for large $N_{u}(t)$, each language is characterized by a scaling
relation
\begin{equation}
\sigma_{r}(t|f_{c}) \sim N_{u}(t|f_{c})^{-\beta} \ ,
\label{eqSV}
\end{equation}
with language-dependent scaling exponent $\beta \approx 0.08 -0.35$.
We use $f_{c} = 10/Min[N_{u}(t)]$, which defines the frequency threshold for the inclusion of a given word in our  analysis.
There are two candidate null models which give insight into the limiting behavior of $\beta$.
The Gibrat proportional growth model predicts $\beta = 0$ and the Yule- Simon urn model predicts $\beta = 1/2$ \cite{SizeVar2}.
We observe $\beta<1/2$, which indicates that the fluctuation scale decreases more slowly with increasing corpora size than   would be expected
from the Yule-Simon urn model prediction, deducible via the ``delta method'' for determining the approximate scaling of a distribution and its standard deviation $\sigma$ \cite{DeltaMethod}.

To further compare the roles of the kernel lexicon versus the unlimited lexicon, we apply our pruning method to quantify the dependence of the scaling exponent $\beta$ on the fluctuations arising from rare words. We omit words from our calculation of $\sigma_{r}(t|U_{c})$ if their use $u_{i}(t)$ in year $t$ falls below the word-use threshold $ U_{c}$.  Fig.~\ref{CorpusTotalSizefcSigmaScaling}(B) shows that $\beta(U_{c})$ increases from values close to 0 to values less than 1/2 as $U_{c}$ increases exponentially. An increasing $\beta(U_{c})$ confirms our conjecture that  rare words are largely responsible for the fluctuations in a language. However, because of the dependency structure between words, there are residual fluctuation spillovers into the kernel lexicon likely accounting for the fact that  $\beta < 1/2$ even when the fluctuations from the unlimited lexicon are removed.

A  size-variance relation showing that larger entities have smaller characteristic fluctuations
was also demonstrated at the scale of individual words  using the same {\it Google n-gram}
dataset \cite{petersen_sr12}.
Moreover, this size-variance relation is strikingly analogous to
the decreasing growth rate volatility observed as complex economic entities
(i.e.~firms or countries) increase in size \cite{Growth1b, Growth2,SizeVar1,SizeVar2,SizeVar3,SizeVar4}, which
strengthens the analogy of language as a complex ecosystem of words governed by competitive forces.

Further possible explanations for $\beta <1/2$ is that language growth is counteracted by the
influx of new words which tend to have growth-spurts around 30-50 years
following their birth in the written corpora \cite{petersen_sr12}.
Moreover, the fluctuation scale $\sigma_{r}(t|f_{c})$ is positively influenced
by adverse conditions such as wars and revolutions, since a decrease in $N_{u}(t)$ may decrease the competitive advantage that old words have over new words, allowing new words to
break through. The globalization effect, manifesting from increased human mobility during periods of conflict, is also responsible for the emergence of new words within a language.

\section*{Discussion}
A coevolutionary description of language and culture requires many
factors and much consideration \cite{Mufwene1,Mufwene2}. While
scientific and technological advances are largely responsible for
written language growth as well as the birth of many new words
\cite{petersen_sr12}, socio-political factors also play a strong
role.  For example, the sexual revolution of the 1960s triggered the
sudden emergence of the words ``girlfriend'' and ``boyfriend'' in the
English corpora \cite{google}, illustrating the evolving culture of
romantic courting. Such technological and socio-political perturbations
require case-by-case analysis for any deeper understanding, as
demonstrated comprehensively by Michel et al. \cite{michel_s11}.

Here we analyzed the macroscopic properties of written language using
the \textit{Google Books} database \cite{google}. We find that the word
frequency distribution $P(f)$ is characterized by two scaling
regimes.  While frequently used words that constitute the kernel lexicon
follow the Zipf law, the distribution has a less-steep scaling regime
quantifying the rarer words constituting the {\it unlimited
lexicon}.  Our result is robust across languages as well as across other data
subsets, thus extending the validity of the seminal observation by
Ferrer i Cancho and Sol{\'e} \cite{cancho_jql01}, who first reported it
for a large body of English text. The kink in the slope preceding the
entry into the unlimited lexicon is a likely consequence of the limits
of human mental ability that force the individual to optimize the usage
of frequently used words and forget specialized words that are seldom used. This
hypothesis agrees with the ``principle of least effort'' that minimizes
communication noise between speakers (writers) and listeners (readers),
which in turn may lead to the emergence of the Zipf law
\cite{cancho_pnas03}.

Using an extremely large  written corpora that documents the profound expansion of language over centuries,
 we analyzed the dependence of vocabulary growth on corpus growth and
 validate the Heaps law scaling relation given by Eq. \ref{HLaw}.  Furthermore we systematically prune the corpora data using a word occurrence threshold $U_{c}$, and comparing the resulting $b(U_{c})$ value to the
$\zeta\approx 1$ value, which is stable since it is derived from the ``kernel'' lexicon. We  conditionally confirm  the
theoretical prediction $\zeta \approx 1/b$ \cite{Mandelbrot61,Karlin67,PowerLawOccupancyHeaps,DerivHeaps,serrano_pone09,FiniteSizeHeaps}, which we validate only  in the case that the extremely rare ``unlimited'' lexicon words are not included in the data sample (see Figs. \ref{MultifcEng} and \ref{MultifcOther}).

The economies of scale ($b<1$)  indicate that there is an \textit{increasing marginal return} for new
  words, or alternatively, a \textit{decreasing marginal need} for new
  words, as evidenced by allometric scaling.  This can intuitively be
  understood in terms of the increasing complexities and combinations of
  words that become available as more words are added to a language,
  lessening the need for lexical expansion.
However, a relationship between new words and existing words is retained.
  Every introduction of a
  word, from an informal setting (e.g.~an expository text) to a formal
  setting (e.g.~a dictionary) is yet another chance for the more common
  describing words to play out their respective frequencies,
  underscoring the hierarchy of words.  This can be demonstrated quite
  instructively from Eq.~(\ref{marginal}) which implies that for $b=1/2$
  that $\frac{\partial N_{u}}{\partial N_{w}} \propto N_{w}$, meaning
  that it requires a quantity proportional to the vocabulary size
  $N_{w}$ to introduce a new word, or alternatively, that a
    quantity proportional to $N_{w}$ necessarily results from the addition.

Though new words are needed less and less, the expansion of language continues, doing so with marked characteristics. Taking the growth rate fluctuations of word use to be
a kind of temperature, we note that like an ideal gas, most languages
``cool'' when they expand. The fact that the relationship between the
temperature and corpus volume is a power law, one may, loosely speaking,
liken language growth to the expansion of a gas or the growth of a
company \cite{Growth1b, Growth2, SizeVar1,SizeVar2, SizeVar3,SizeVar4}.
In contrast to the static laws of Zipf and Heaps, we note that this finding is of a dynamical nature.

Other aspects of language growth may also be understood in terms of
expansion of a gas. Since larger literary productivity imposes a
downward trend on growth rate fluctuations --- which also implies that the ranking of the top words and phases becomes more stable \cite{MostCommonWords}  ---productivity itself can be
thought of as a kind of inverse pressure in that highly productive years
are observed to ``cool'' a language off. Also, it is during the
``high-pressure'' low productivity years that new words tend to emerge
more frequently.

Interestingly, the appearance of new words is more like gas
condensation, tending to cancel the cooling brought on by language
expansion.  These two effects, corpus expansion and new word
``condensation,'' therefore act against each other.
Across all corpora we calculate a size-variance scaling exponent $0<\beta<1/2$, bounded by the prediction of $\beta=0$ (Gibrat growth model) and $\beta=1/2$ (Yule-Simon growth model) \cite{SizeVar2}.

In the context
of allometric relations, Bettencourt et al.~\cite{bettencourt_pnas07}
  note that the scaling relations describing the dynamics of
  cities show an {\it increase} in the characteristic pace of life as the
  system size grows, whereas those found in biological systems show {\it
    decrease} in characteristic rates as the system size grows. Since
  the languages we analyzed tend to ``cool'' as they expand, there may be deep-rooted parallels with biological systems based on principles of efficiency \cite{cancho_pnas03}.
  Languages, like biological systems demonstrate economies of scale
  ($b<1$) manifesting from a complex dependency structure that mimics a
  hierarchical ``circulatory system'' required by the organization of
  language
  \cite{WordnetlexiconHierarchy,SemanticNetwork,textHeirarchyCorr,textHeirarchyCorr2, FractalCorrCorpora,ReturnIntervalsLanguage}
  and the limits of the efficiency of the speakers/writers who exchange the words
  \cite{bernhardsson_njp09, altmann_pone11,WordBursts}.

\section*{Acknowledgments}
\noindent  AMP acknowledges support from the IMT Lucca Foundation. JT, SH and HES acknowledge support from the DTRA, ONR, the European EPIWORK and LINC projects, and the Israel Science Foundation. MP acknowledges support from the Slovenian Research Agency.\\

\section*{Author Contributions}
\noindent  A. M. P., J. T., S. H., H. E. S., \& MP designed research, performed research, wrote, reviewed and approved the manuscript. A. M. P.  performed the numerical and statistical analysis of the data.
\end{document}